\documentclass[a4paper]{jpconf}
\usepackage{graphicx}
\usepackage{url}
\begin{document}
\title{The Capabilities of the upgraded  MIPP experiment
with respect to HyperNuclear Physics}

\author{Rajendran Raja\\
For the MIPP collaboration}

\address{Fermilab, P.O. Box 500, Batavia, Illinois 60510}

\ead{raja@fnal.gov}

\begin{abstract}
We describe the state of analysis of the MIPP experiment, its plans 
to upgrade the experiment and the impact 
such an upgraded experiment will have on
hypernuclear physics.
\end{abstract}

\section{Introduction}
\begin{figure}[h]
\centerline{\includegraphics[width=0.6\textwidth]{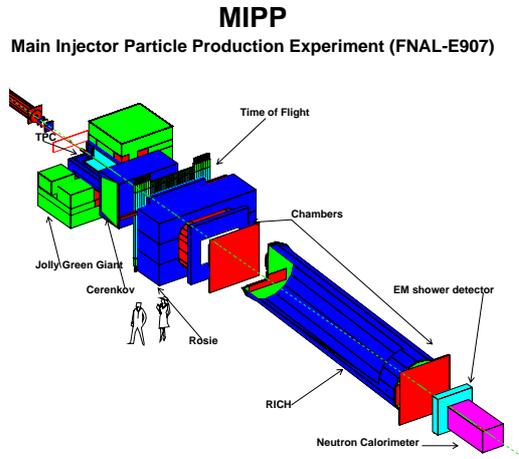}}
\caption{Layout of the experiment - Run I.  With the MIPP upgrade we propose to 
add the plastic ball detector as a recoil detector surrounding the target.}
\label{layout}
\end{figure}

The upgraded MIPP experiment is designed to measure the properties of strong
interaction spectra form beams $\pi^\pm, K^\pm$ and $p^\pm$ for momenta
ranging from 1 GeV/c to 120 GeV/c.   The
layout of the apparatus in the data taken so far can be seen in
Figure~\ref{layout}. The centerpiece of the experiment is the time
projection chamber, which is followed by the time of flight counter, a
multi-cell Cerenkov detector and the RICH detector. The TPC can
identify charged particles with momenta less than 1~GeV/c using dE/dx,
the time of flight will identify particles below approximately 2
GeV/c, the multi-cell Cerenkov detector is operational from ~2.5 GeV/c
to 14~GeV/c and the RICH
detector can identify particles up to 120 GeV\/c.  Following this is an 
EM and hadronic calorimeter capable of detecting forward going neutrons 
and photons.

\begin{figure}[h]
\centerline{\includegraphics[width=0.6\textwidth]{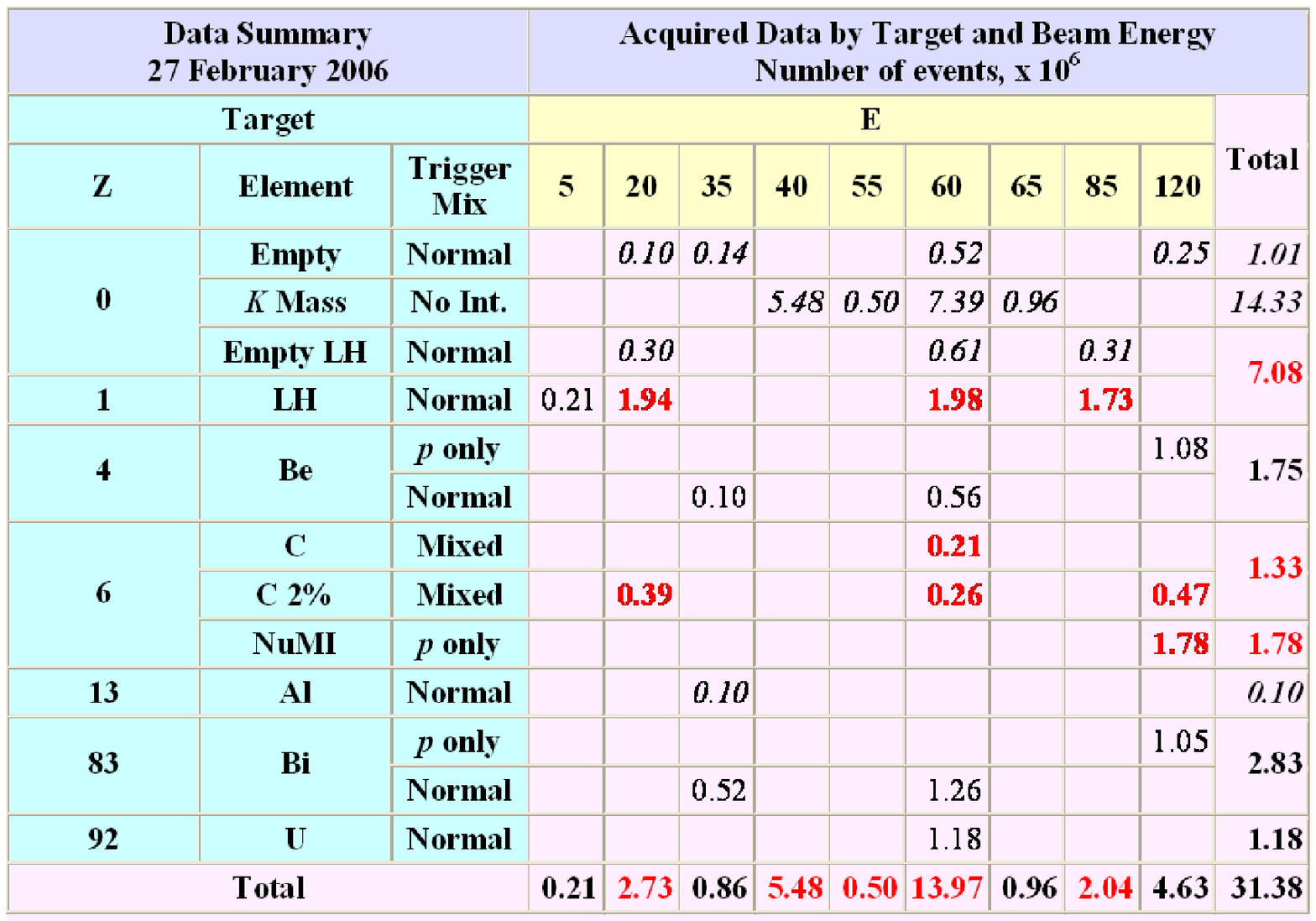}}
\caption{Data taken  by MIPP I to date}
\label{runtable}
\end{figure}

The experiment has been busy
analyzing its data taken on various nuclei and beam conditions. The
table~\ref{runtable} shows the data taken by MIPP I to date.

\begin{figure}[htb!]
\begin{minipage}{30pc}
\includegraphics[width=\textwidth]{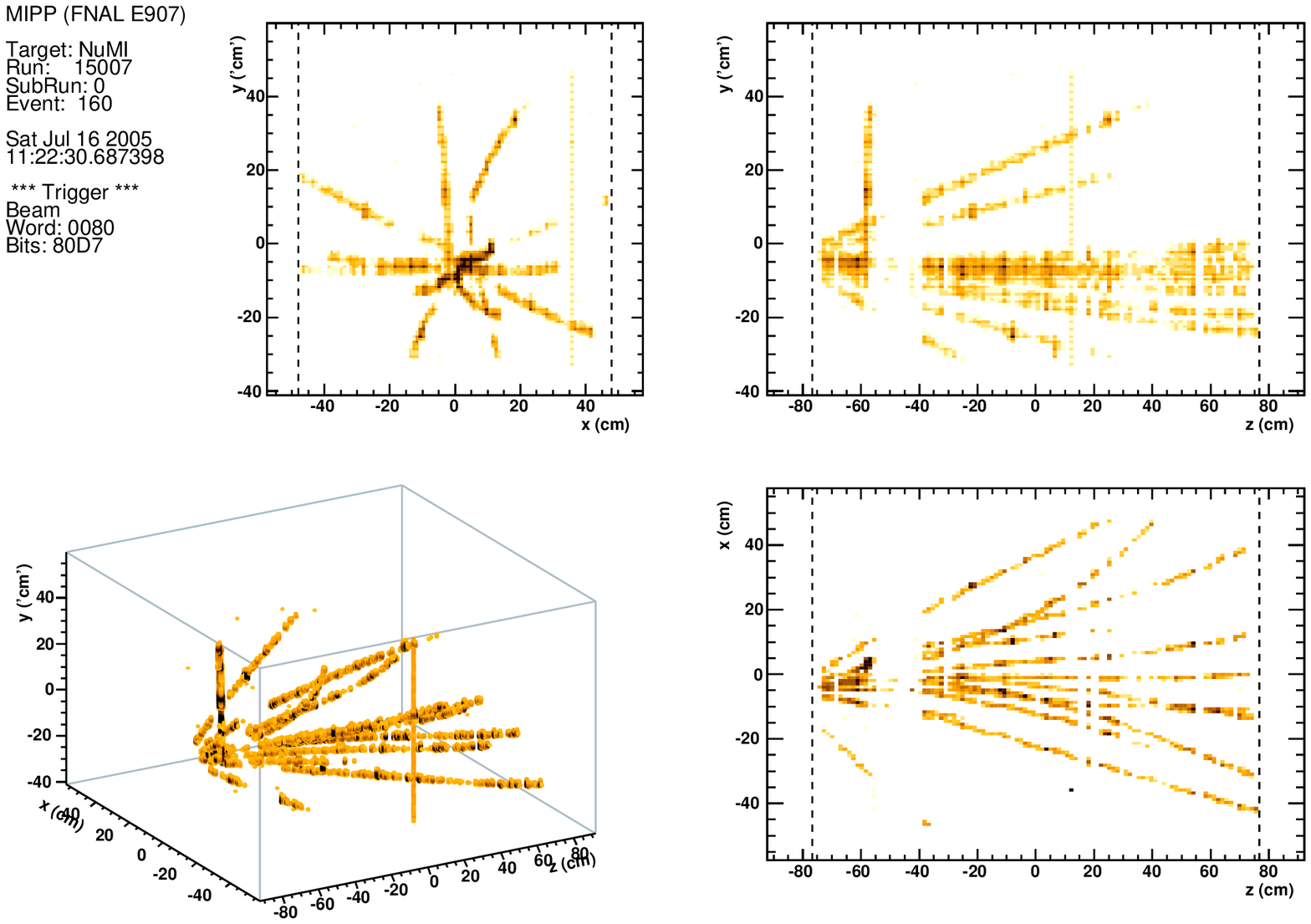}
\end{minipage}
\begin{minipage}{30pc}
\includegraphics[width=\textwidth]{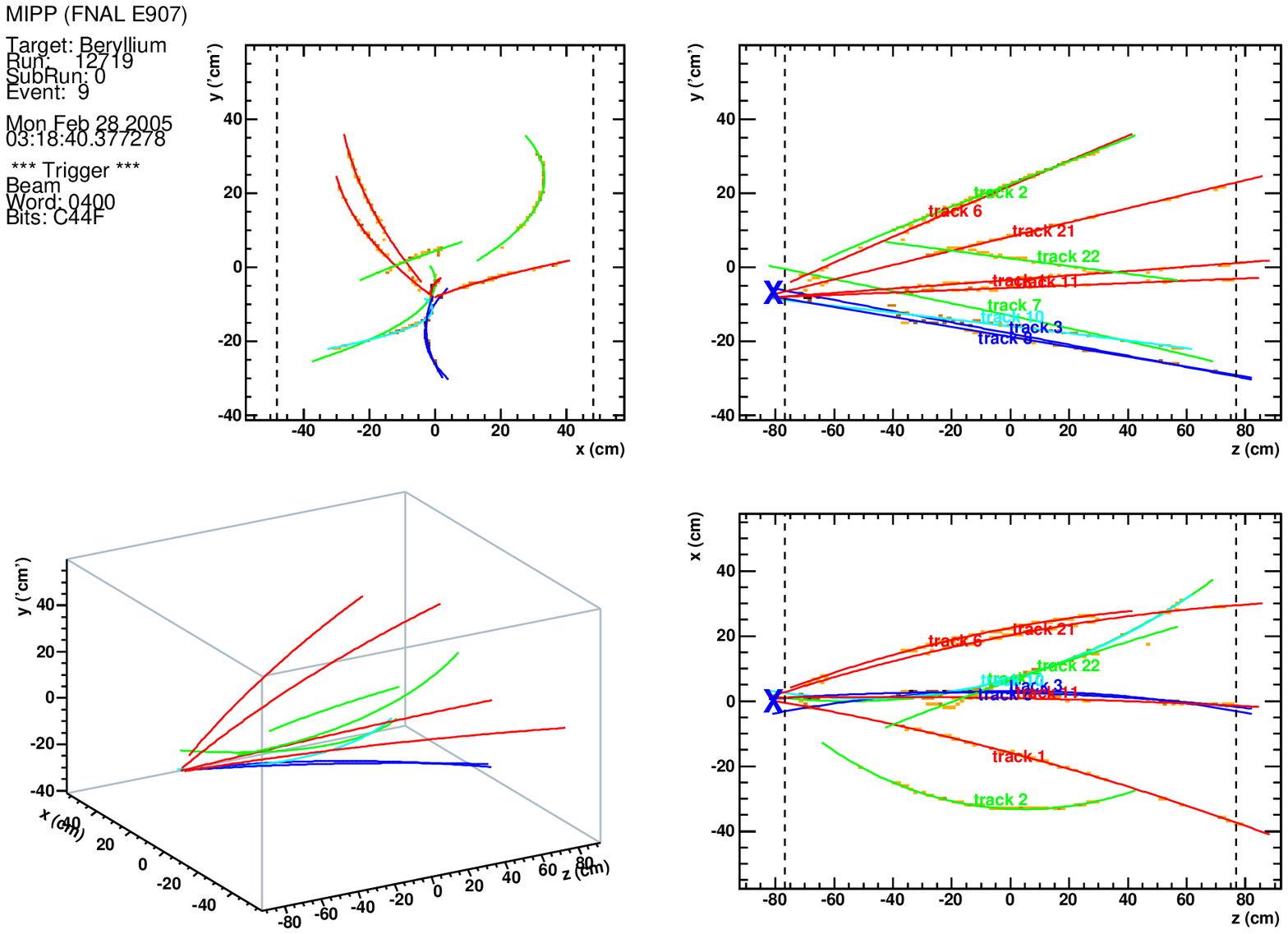}
\end{minipage}
\caption{RAW and Reconstructed TPC tracks  from two different events.}
\label{tpc}
\end{figure}

We have almost complete acceptance in the forward hemisphere in the
lab using the TPC. The reconstruction capabilities of the TPC can be seen in Figure~\ref{tpc}.

\begin{figure}[h]
\centerline{\includegraphics[width=0.6\textwidth]{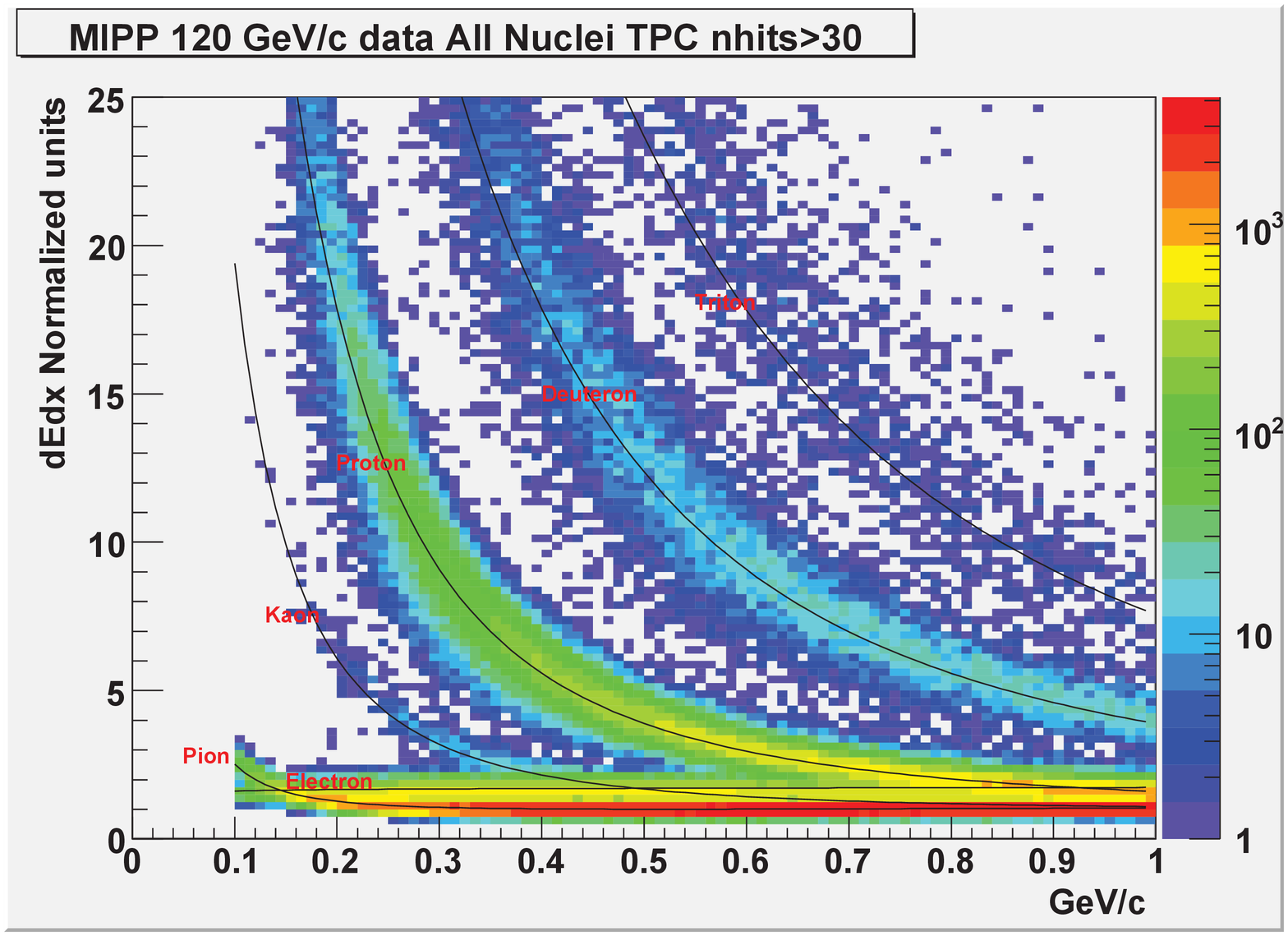}}
\caption{TPC dE/dx data from Run I. Electrons, pions, kaons and protons 
are distinguishable. Deuterons and tritons can also be seen}
\label{tpcdedx}
\end{figure}

The particle identification capabilities of the TPC can be seen in Figure~\ref{tpcdedx}.

\begin{figure}[h]
\centerline{\includegraphics[width=0.6\textwidth]{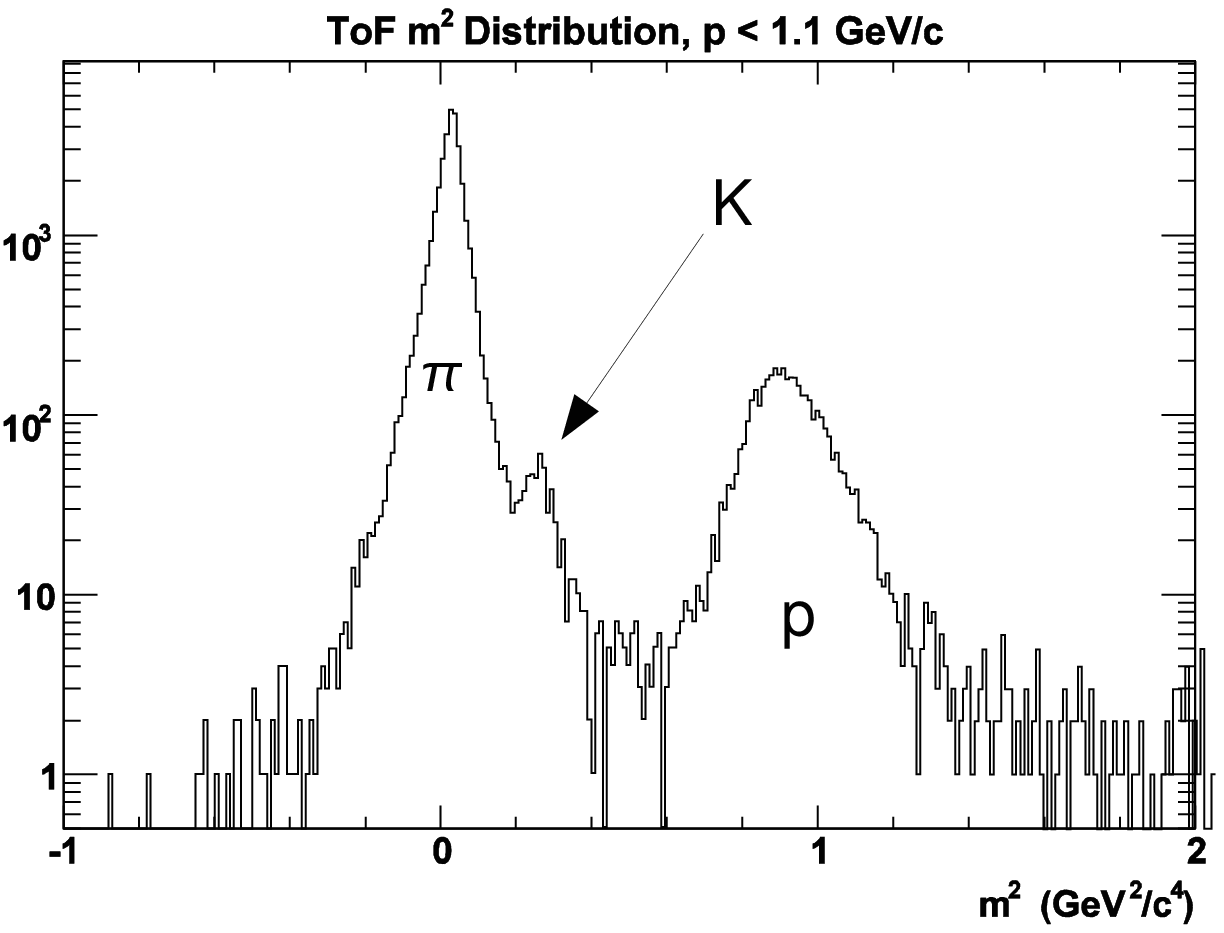}}
\caption{ToF data with momenta less than 1.1 GeV/c. 
a kaon peak is clearly visible.}
\label{tof}
\end{figure}

The time of flight system provides further measurement of the particles with 
momenta less than 2~GeV/c.  Figure~\ref{tof} shows the time of flight data 
where a kaon peak is clearly visible.

\section{Results from data taken so far}

MIPP has published forward going neutron spectra off various nuclei and compared it
with Monte Carlo predictions~\cite{neutron}

\begin{figure}[h]
\centerline{\includegraphics[width=0.6\textwidth]{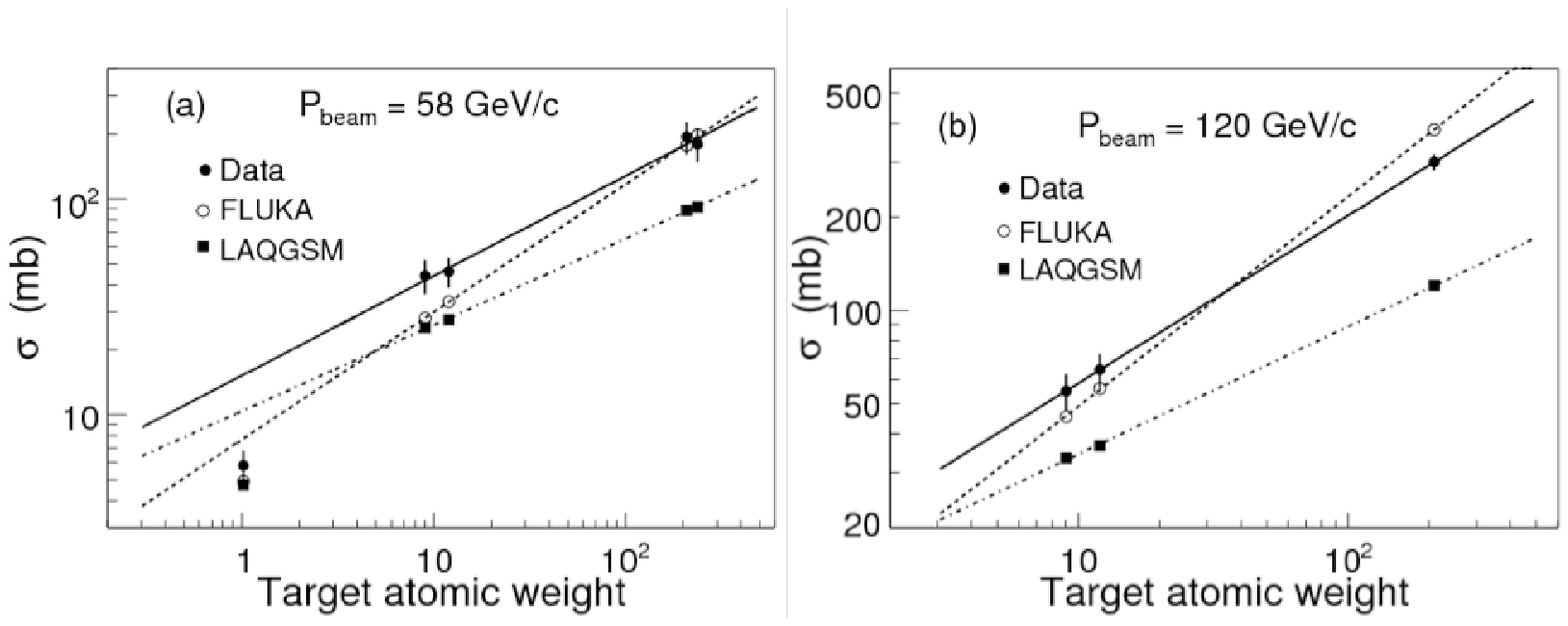}}
\caption{Comparison of the A-dependence of MIPP cross sections with
those from Monte Carlos. The lines are fits to the data. The cross
sections are for producing neutrons with momentum greater than the
threshold and within an angular range of 20.4 mrad. The errors are
combined statistical and systematic uncertainties. Note that the
hydrogen data point is not included in the fit. The cross sections are
not corrected for geometric acceptance.}
\label{neutrons}
\end{figure}
It can be seen clearly that the Monte Carlo predictions differ significantly 
from each other and the data.

\begin{figure}[h]
\centerline{\includegraphics[width=0.6\textwidth]{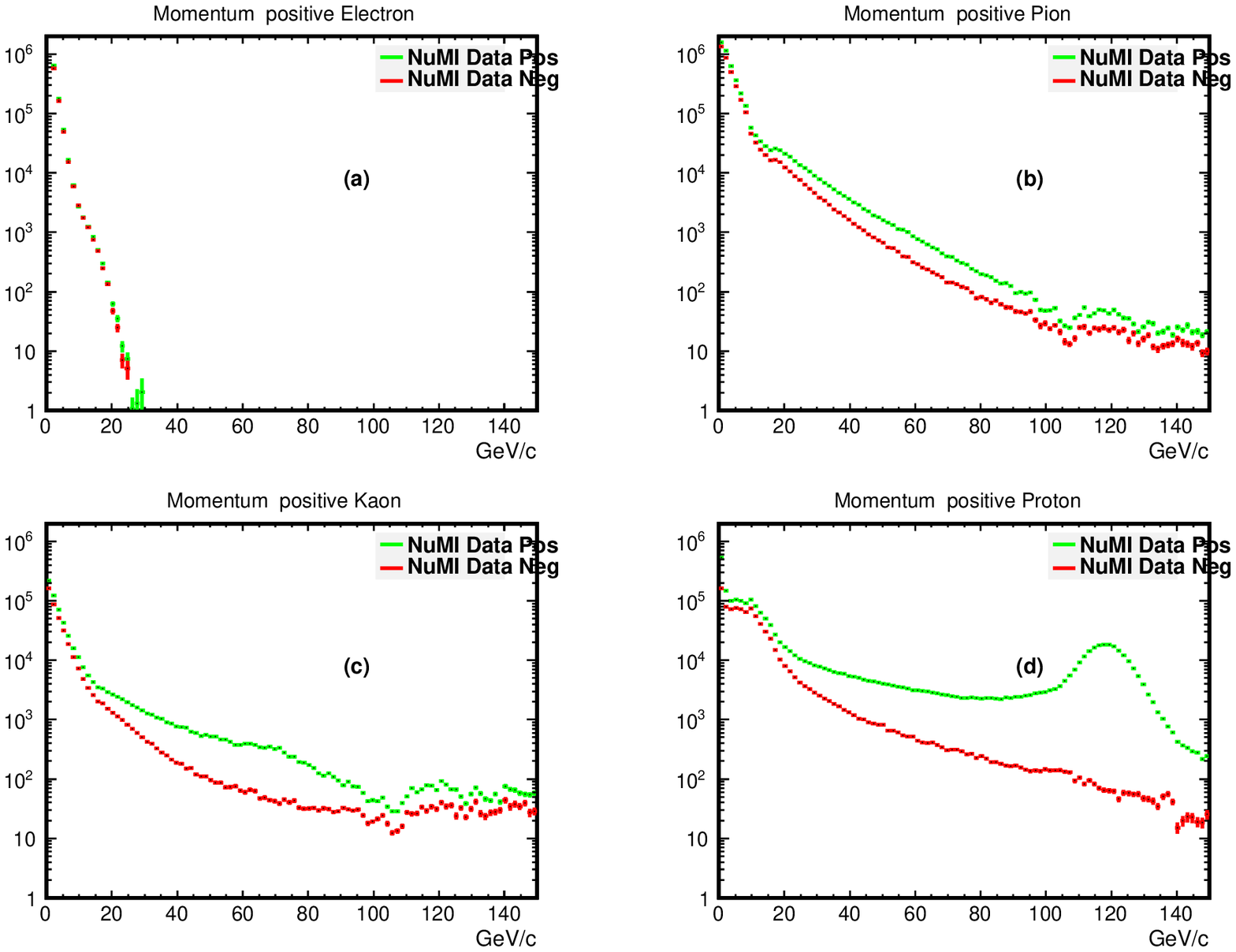}}
\caption{Analysis of NuMI target spectra (data) as a function of particle type (preliminary)}
\label{numidata}
\end{figure}

Figure~\ref{numidata} shows (in a paper in preparation,~\cite{numi}) 
preliminary results of particle 
production from the full NuMI target as a function of momenta. 

\section{The MIPP Upgrade proposal}

\begin{figure}[h]
\centerline{\includegraphics[width=0.6\textwidth]{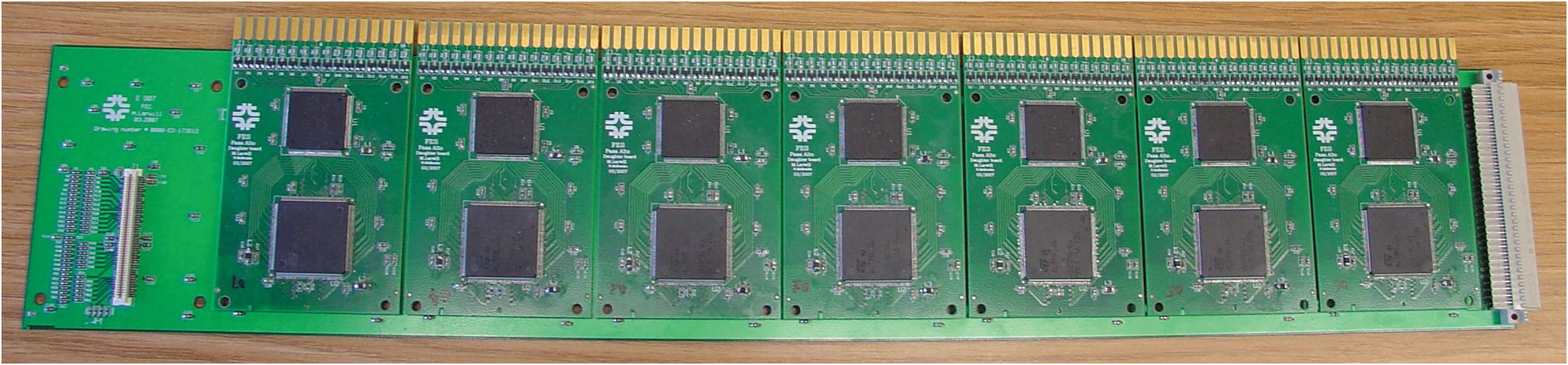}}
\caption{Electronics prototype of TPC upgrade board with ALTO/PASA chips}
\label{tpcchips}
\end{figure}

The MIPP upgrade proposal consists primarily of an electronics upgrade
of the whole detector so that the detector reads out at 3~kiloHZ as
compared to 20~Hz in Run I. All the new electronics modules have been
prototyped. The TPC has the largest data rate and we display one of its
elctronics 
modules in Figure~\ref{tpcchips}. 
The document~\cite{upgrade_new}
contains the details of the upgrade proposal. The MIPP upgrade
proposal will acquire high quality data of particle production with
high statistics and much improved particle identification. The
available beam momenta and the calculated beam rates 
can be seen in table~\ref{mippbeam} and range approximately 1~GeV/c to 120
GeV/c.
As such the upgrade can straddle the resonance region and the
scaling region of strong interaction physics using 6 beam species. 
The MIPP upgrade intends
to acquire data for a variety of interactions. These include
\begin{itemize}
\item A systematic study of missing baryon resonances between 1~GeV and 3~GeV.
\item An investigation of fundamental scaling laws in inclusive reactions.
\item A thorough investigation of forward neutron spectra.
\item A systematic acquisition of particle production data and comparison 
and tuning of existing Monte Carlo generators.
\item A measurement of particle production data from full scale neutrino 
targets and using the data to improve the target design.
\item A systematic study of hypernuclei
\item Measurement of cross section off nitrogen targets which is essential
to understanding the cosmic ray showers and atmospheric neutrino production for
experiments such as INO.
\end{itemize}
This is made possible since
MIPP upgrade will be able to collect
5~millions events per day assuming a beam spill rate of one 4
second spill every 2 minutes,

\begin{table}[h]
\begin{tabular}{|c|c|c|c|c|c|c|c|}
\hline
\hline
momentum & p & K$^+$ & $\pi^+$ & momentum & $\bar{p}$ & K $^-$ & $\pi^-$ \\
GeV/c    &   &       &         & GeV/c    &           &        &         \\
\hline
\hline
1 &  5752 &  0 &  64798 &  1 &  7907 &  0 &  30425 \\
\hline
2 &  23373 &  194 &  459718 &  2 &  26863 &  142 &  236494 \\
\hline
3 &  53431 &  3060 &  1069523 &  3 &  51424 &  2221 &  598742 \\
\hline
5 &  153220 &  32763 &  2400799 &  5 &  103996 &  23164 &  1550810 \\
\hline
10 &  663916 &  223210 &  5006708 &  10 &  195767 &  142777 &  3862225 \\
\hline
15 & 1618120 &  443557 &  7141481 &  15 &  221602 &  245868 &  5248463 \\
\hline
20 &  3113387 &  655426 &  9290219 &  20 &  212171 &  306685 &  5841030 \\
\hline
30 &  8158054 &  1043430 &  12770579 &  30 &  160329 &  340144 &  5837467 \\
\hline
40 &  16664431 &  1294189 &  13944272 &  40 &  101617 &  288728 &  5156862 \\
\hline
50 &  29288928 &  1338452 &  12788523 &  50 &  53056 &  196400 &  4114582 \\
\hline
60 &  45985629 &  1191744 &  10094311 &  60 &  22092 &  108032 &  2905091 \\
\hline
70 &  65227010 &  919279 &  6834097 &  70 &  6987 &  47093 &  1762060 \\
\hline
\hline
\end{tabular}
\caption{Calculated Beam rates for $2\times10^{11}$ protons on primary target
for MIPP upgrade\label{mippbeam}}
\end{table}

\section{Hypernuclear Physics with MIPP Upgrade}

\begin{figure}[h]
\centerline{\includegraphics[width=0.6\textwidth]{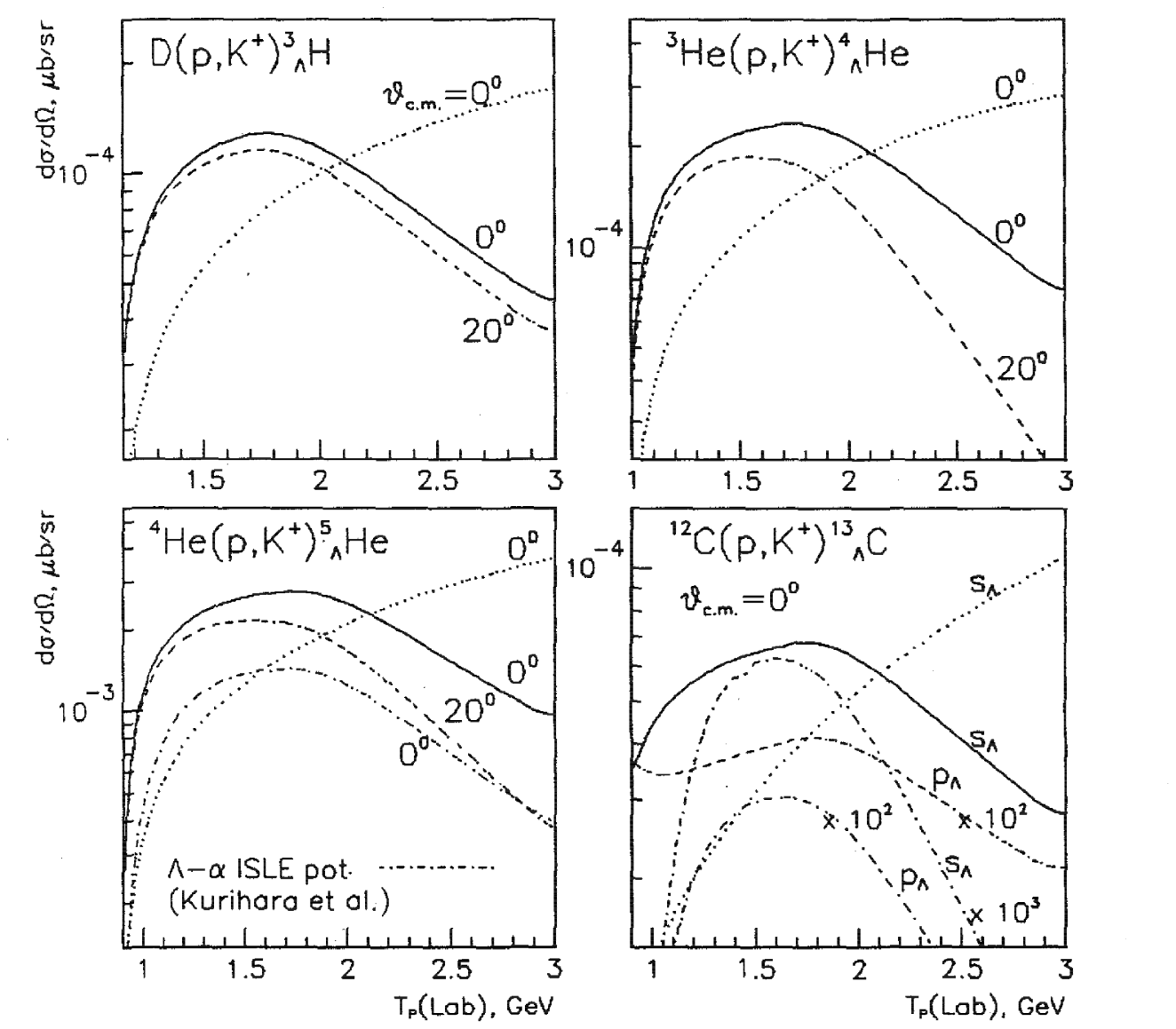}}
\caption{Differential cross sections of single hyper nucleus production on
various nucleus with a proton beam and a final state kaon~\cite{fetisov}}
\label{single}
\end{figure}

\begin{figure}[h]
\centerline{\includegraphics[width=0.6\textwidth]{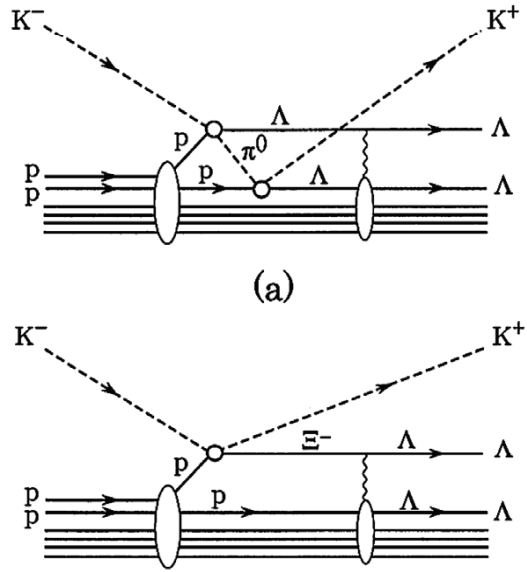}}
\caption{Mechanisms for producing double hypernuclei using a $K^-$ beam.
$K^+$ beams can study the background to this process~\cite{bhang}}
\label{double}
\end{figure}

\begin{figure}[h]
\centerline{\includegraphics[width=0.6\textwidth]{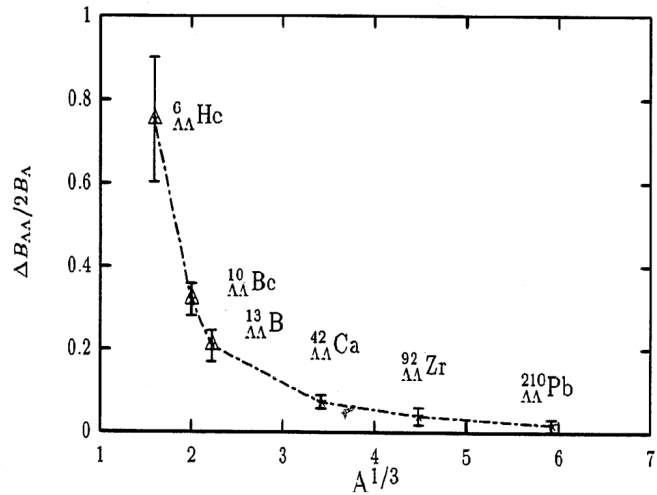}}
\caption{Binding energy of hypernuclei as a function of atomic weight.
\label{binding}}
\end{figure}

\begin{figure}[h]
\centerline{\includegraphics[width=0.6\textwidth]{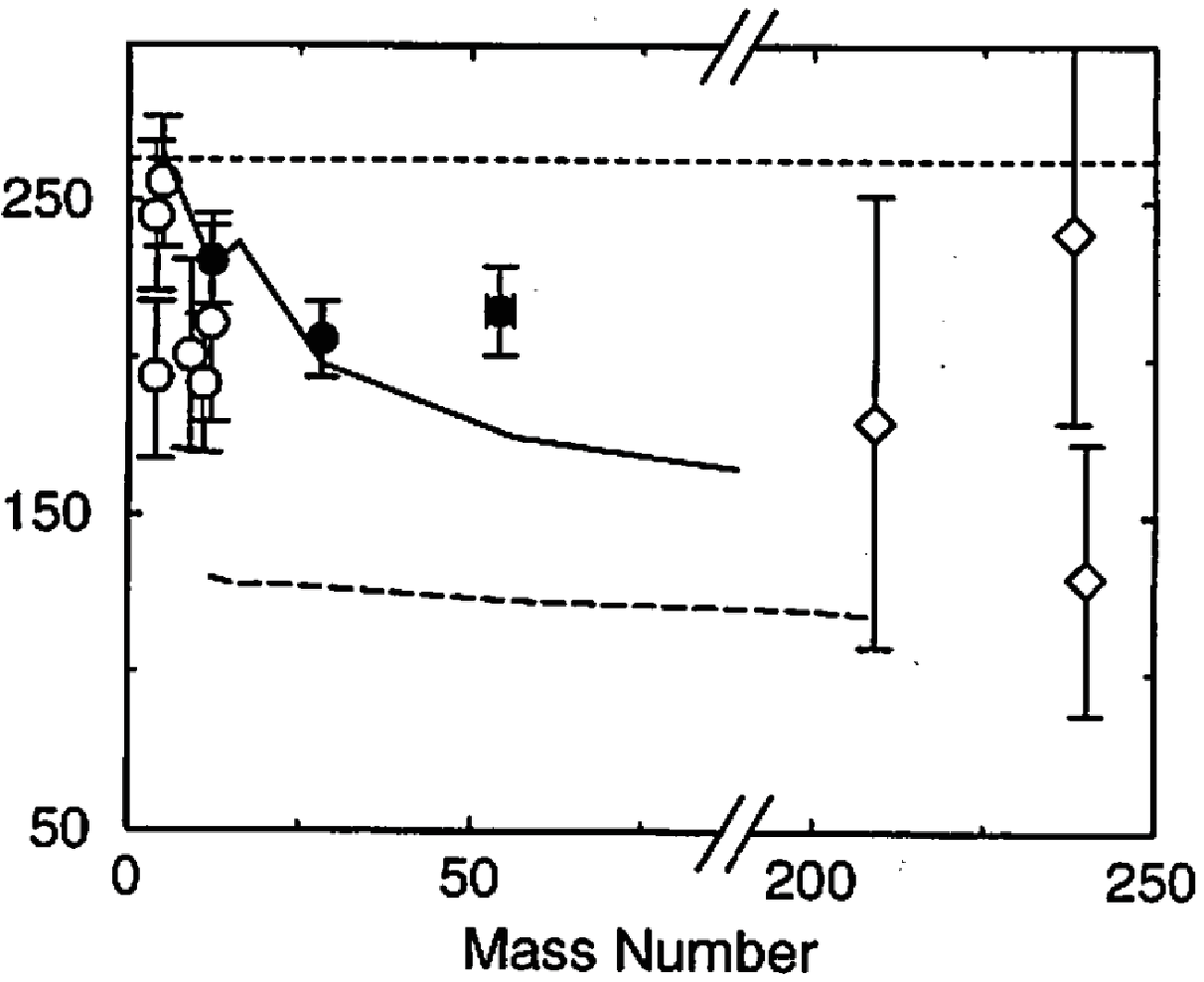}}
\caption{Lifetimes of hypernuclei as a function of atomic weight A. As A
increases, lifetimes get shorter, implying less tightly bound hyper nuclei}
\label{lifetimes}
\end{figure}

Single strange hypernuclei (with one $\Lambda$ or one $\Sigma$) have
production cross sections in the 100 to 250 nbarn/steradian at 1 to 2
GeV/c proton beams on nuclear targets as can be seen for
Figure~\ref{single}. Doubly strange hypernuclei (with two $\Lambda$)
are 10 to 50 nbarn/sr with a kaon or hyperon beam on nuclear target and
are more difficult to measure.  The diagrams for doubly strange
hypernuclei can be found in Figure~\ref{double}. Doubly strange
hypernuclei with one $\Xi$ are not discovered yet, but 7 candidate
events are known to exist.
The binding energies for nuclei as a function of atomic weight can be
found in Figure~\ref{binding}. The binding energies decrease with
atomic weight indicating that the hypernuclei are less tightly bound as
atomic weight increases. This behavior is also reflected in the
lifetimes plot in Figure~\ref{lifetimes} where the lifetimes of the
bound lambda is seen to decrease with increasing atomic weight.

Using MIPP upgrade, we have complete forward acceptance using the TPC
and full particle identification (using the TPC and the time of flight)
for particles of momentum less than 1.5~GeV/c. In the backward hemisphere in the lab frame, 
we have the plastic ball that catches recoil particles.

We investigate the channels that are likely to yield hypernuclei.
 We employ the high energy  physics notation to denote the reactions.

The channels involving a $\pi^+$ beam can be denoted by
\begin{equation}
\pi^+ + p \rightarrow \pi^+ + K^+ + \Lambda
\end{equation}
In the above equation, $\pi^+$ beam remains a $\pi^+$ beam and fragments
a proton target to form a $K^+$ and a $\Lambda$.  Such reactions are called 
diffractive dissociations
of the target and these maintain a large cross section as the beam energy increases.
The $K^+$ in the
final state is detected by the TPC and the time of flight systems
 and the $\Lambda$ is absorbed in the nucleus to
make the hypernucleus. There may be other particles in the finally
state, which the particle identification of MIPP will ensure contain a
net strangeness of zero. The counterpart of the above reaction with a 
neutron target can be written
\begin{equation}
\pi^+ + n \rightarrow \pi^+ + K^0_S + \Lambda
\end{equation}
The $K^0_s$ in the final state can be detected by the TPC.
Similarly with a proton beam
\begin{equation}
p + p \rightarrow p + K^+ + \Lambda 
\end{equation}
\begin{equation}
p + n \rightarrow p + K^0_S + \Lambda
\end{equation}
These form the proton counter part.
Using a K$^-$ beam,
one can look for the reaction
\begin{equation}
K^- + p \rightarrow \pi^+ +   \Lambda 
\end{equation}
\begin{equation}
K^- + n \rightarrow   \pi^0 + \Lambda
\end{equation}
One can test the backgrounds to this reaction using K$^+$ beams  which should 
not produce hypernuclei.
\section{Conclusions}
The high acceeptance and particle identification of MIPP upgrade promises to produce
a high quality hypernucleus experiment.
This work was done at Fermilab operated by Fermi Research Alliance, LLC 
under Contract No. De-AC02-07CH11359 
with the United States Department of Eneegy.  

\end{document}